# Leveraging a Relationship with Biology to Expand a Relationship with Physics

Vashti Sawtelle[1,2] & Chandra Turpen[3]
[1]Lyman Briggs College and [2]Department of Physics and Astronomy, Michigan State University, East Lansing, MI, 48825; [3]Department of Physics, University of Maryland, College Park, MD 20742

This work examines how experiences in one disciplinary domain (biology) can impact the relationship a student builds with another domain (physics). We present a model for disciplinary relationships using the constructs of identity, affect, and epistemology. With these constructs we examine an ethnographic case study of a student who experienced a significant shift in her relationship with physics. We describe how this shift demonstrates (1) a stronger identification with physics, (2) a more mixed affective stance towards physics, and (3) more expert-like ways of knowing in physics. We argue that recruiting the student's relationship with biology into experiences of learning physics impacted her relationship with physics as well as her sense of how physics and biology are linked.

## I. INTRODUCTION

Science education is undergoing a fundamental shift toward increasing interdisciplinarity. In physics, this movement has manifested as an increasing focus on the intersection of physics and biology. Recently the community has seen evidence of this movement in special editions of *CBE – Life Sciences Education* and the *Am. J. Phys.*, a Gordon Research Conference, and a conference hosted by AAPT on Introductory Physics for the Life Sciences. While attending one of these venues for discussing teaching physics to life science students, we encountered a number of ways that physics instructors positioned biology students, such as seeing these students as:

- Used to memorizing facts in order to succeed
- Not liking equations and mathematics and/or "scared to death" of math
- Primarily interested in health applications
- Not prepared in physics (i.e. unlikely to have taken physics in high school)
- Not prepared in math
- Highly motivated by grades, and willing to work hard to achieve them
- Having an aversion to mathematics, but appreciating pictures

This paper describes the development of an introductory physics case-study student whose incoming tendencies demonstrate consistency with many of these stereotypes. We provide a way to begin to engage with these stereotypes in an empirical, rather than intuitive manner. In this paper we challenge these stereotypes as immutable characterizations of individual students. We explore the case of a student whose relationship with physics expands throughout a physics course designed to leverage biology experiences. We posit that leveraging the student's affect and identity toward biology explains this expanding relationship with physics. From an instructional standpoint, this work demonstrates that the expectations with which our students enter our classrooms should not be treated as static or immutable characteristics. Instead, we argue that instructors can and should proactively interact with and try to mold these expectations.

## II. A CASE OF AN EXPANDING RELATIONSHIP WITH PHYSICS

This paper describes an account of how a student who described herself as resistant to physics opened up and began participating in physics in productive ways. This case has been developed around a longitudinal study of "Violet," a student enrolled in a university-level Introductory Physics for Biologists course (NEXUS/Physics). The course was transformed to build coherence between physics and biology [1-4]. Violet initially caught our attention because she was distraught during the first recitation of the course, nearly in tears and considering dropping the course. The teaching assistant (TA) attributed Violet's response to her insecurity with doing the math in the first group problem-solving activity. The following day in class, a participant observer heard her asking for "someone smart" to sit in her group because she is a "flunk" at physics. We were initially interested in understanding how Violet's sense of competence in physics developed throughout the course, and began to follow her more closely. This case study involved in-depth ethnographic participant observations and field notes of in-class discussions in lecture, regular video recordings of Violet's group in lecture and recitation, and periodic individual interviews of Violet with the first author throughout the academic year.

As we began to analyze the data we noticed that Violet's development was not only in her feelings of competence in physics. Her affect around physics, her sense of what physics is, and her identification with physics were all expanding, and we began conceptualizing these constructs as comprising her *relationship* with physics. We begin this work by articulating the theoretical frameworks we are bringing to the idea of *relationship* and expansion of a relationship. We then carefully document Violet's incoming and outgoing relationships with physics. We follow this careful description by discussing how the constructs of identity, affect, and epistemology help us to explain the shift. Finally, we return to our original motivation for this study by discussing how these constructs provide an understanding of the access points for interacting with stereotypes surrounding students in physics classrooms. Based on our analysis, we see Violet's relationship with physics changing from someone who "hates physics", is "not a symbols person", and thinks "physics just happens" to a person who has a "love/hate" relationship with physics, is "more of a symbols person", and believes physics pursues "many pathways for understanding equations." These changes are generative for understanding how Violet sees physics and biology as related. This account has implications for thinking about how expertise in (or relationships with) one arena may be leveraged to impact affinity towards (and expertise in) other domains.

III. THEORETICAL COMMITMENTS: MODELING DISCIPLINARY RELATIONSHIPS THROUGH IDENTITY, AFFECT, & EPISTEMOLOGY

We begin modeling Violet's change by conceptualizing a relationship with a discipline. Using the term "relationship" arises primarily from Violet's own account of the change over the year-long course, but has affordances in modeling the change for Violet. Building a relationship requires interaction between two parties, and thus a relationship with the discipline focuses our attention on the role both the discipline and the student play in constructing the relationship. In modeling the change Violet experiences, we attend to three interacting components: identity, affect, and epistemology. We view these constructs as interacting and influencing each other, as represented by Figure 1. A student's sense of who they are and how they fit into various social situations in a moment and across moments (identity) is tied up with the emotions they experience (affect). In a given classroom context, both sense of self and emotions are dependent on and influence the student's views about what it means to know and learn in a particular discipline (epistemology). In the following section we elaborate on our definitions of identity, affect, and epistemology and how they are interrelated.

- Identity: When a person acts or interacts with others we describe him/her as acting like a certain "kind" of person. Gee [5] described this recognition (by oneself and by others) as demonstrating a particular identity. Across theoretical orientations identity has been leveraged to describe engagement and participation in the disciplines [6-8]. In our work, we think of identity as an active practice, rather than a static view on the world [9]. This view takes the perspective that identities are "lived in and through activity" (p.5) rather than stable beliefs that we hold within ourselves, and so must be conceptualized as they develop in the social practice.

- Affect: The affective domain ranges widely from issues of interest and motivation to emotions and goal orientation [10,11]. Outside of test anxiety and attribution theory, work on the role of affect in education is a relatively new arena [11]. We primarily draw on Zembylas [12] who describes affect as performative act, or a focus not on what emotions mean but what they do in a given context.
- Epistemology: Finally, epistemology has a long history in the physics education community as beliefs about knowing and learning [13-16]. Watkins and Elby [16] describe epistemological stances as incorporating both views of nature and knowledge and the kinds of knowing and learning that are rewarded in courses. In this paper we will use epistemology primarily as a way of understanding what kinds of knowing and learning "count" within a particular discipline.

While the lines of research on affect, identity, and epistemology have been productive, they have largely developed independently, though emerging research is beginning to examine the ways they interact (discussed below).

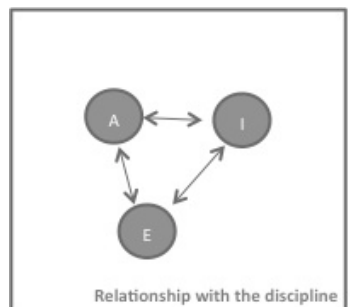

FIG. 1. A disciplinary relationship conceptualized as interactions between affect, identity, and epistemology. The boundary line represents the extent to a single disciple is represented by these interactions.

We see the literature on identity coupling to work on affect through an attention to feelings and interests particularly as they relate to activities in the disciplines [17-19]. Gee and Green [19] clearly link identity to affect when they define socially situated identity as "assembling situated meanings about what identities are relevant to the interaction, with concomitant attitudes and ways of feeling, ways of knowing and believing, as well as ways of acting and interacting" (p.139). The attention to feelings and attitudes in the development of identity makes it clear that, for these authors, affect is tightly coupled with identity.

We treat affect as also tightly coupled with ways of knowing and learning, or epistemology. Gupta, Danielak, and Elby [20] have demonstrated that even at small time scales we can see emotion influencing and sustaining ways of thinking in a classroom. While other works on epistemology have not centrally foregrounded the link with emotions in their accounts, we can see affect and emotions playing a role in discussions of salient episodes. Warren and Rosebery [21] and Engle and Conant [7] do not explicitly speak to issues of affect in their work, but it is clear from their writing that emotions play an important role in providing evidence of engagement in or the sustaining of the scientific epistemology of the classroom. Recent work from Jaber [22,23] has also documented ways that affect is inherent in student experiences of disciplinary practices. For example, Jaber describes an elementary student's experience with magnetism where the student (Sandra) describes getting interested in magnetism in fourth grade "when we actually worked with it, it just FASCinated me and I wanted to learn MORE…" (p. 55). Jaber elaborates on this example and the preceding episode of Sandra working with magnetism suggesting we recognize "Sandra's delight as an emergence of *epistemic affect*" (p. 55). While other authors ([7] and [21]), do not distinguish between emotions as part of the disciplinary engagement and as indicators of engagement, we take the perspective that aligns with Jaber that affect is a central part of the construct we are investigating (which explicitly foregrounds the role of affect in launching and sustaining students' scientific pursuits).

    Connecting epistemological beliefs to developing identity is slightly more established in the literature. In understanding what contributes to the retention and persistence of an engineering student, Danielak, Gupta, and Elby [24] describe how "Michael's" personal epistemological stance toward making sense of things is entangled with his identity and evaluation of belonging in an engineering major. The authors argue that understanding Michael's sense of self relies on understanding his epistemological stance. Other work on identity [7,19,25,26] has suggested that this construct is tightly coupled with the activities being performed, and the specific epistemological criteria being brought to bear [27].

    From these works we can plausibly see how affect is tightly coupled with epistemological stances in science classrooms, as well as with issues of developing identity. In our work as we characterize a relationship with a discipline, we carefully attend to these individual constructs, but also to the ways they interact with one another. Our view of a relationship with a discipline focuses on characterizing the ways these constructs and interactions develop and change throughout time.

### A. Drawing from expansive framing

In our work to characterize Violet's relationship with the discipline, we draw from Engle's work on expansive framing and its connection to transfer [28,29]. In Engle's work [29] framing is broadly defined as the "metacommunicative act of characterizing what is happening in a given context and how different people are participating in it (p.217)." In Engle's model of transfer framing has implications for moving practices from one setting to another: *bounded framing* lacks a sense of connection between two learning environments, while *expansive framing* involves strong links between what is learned in one environment and its relevance to another.

In our study of Violet's changing relationship with the disciplines, we use the term expanding to align with Engle's model of expansive framing. A bounded relationship is one that has a clear limit to the realm of applicability (see the solid line in Figure 1). In this work this boundary is linked to the disciplinary realm of applicability. A relationship that expands then is one that weakens or extends this boundary. As we will show in the case of Violet, an expanding relationship may be one that has some linkages with other disciplinary relationships or one with less distinct boundaries.

## IV. UNDERSTANDING VIOLET'S EXPANDING RELATIONSHIP THROUGH ETHNOGRAPHIC METHODS

This work takes as its central focus a single student's expanding relationship with physics. Our intention in beginning this study was to capture the various elements of practice and experience that shape a student's view of and participation in an introductory physics class. In part we were driven by past experiences with an earlier iteration of the introductory physics class where a student's sense of capability in physics was strongly impacted by interactions with the professor. We were unable to empirically document that student's developmental trajectory, but it served as motivation to focus our attention on similar cases in future iterations. It was with this backdrop that we entered Violet's class. During the study we were simultaneously following three additional case study students starting before the class began, and when the TA mentioned Violet as someone who was worried about her ability to succeed in this environment we quickly focused our analytic attention on her.

### B. Course context and participant description

The pilot run of the NEXUS/Physics course was implemented with a small cohort of students (N=20) to develop prototype materials and closely observe student reasoning in the 2011-2012 academic year. The following academic year the secondary implementation of the course ran with two small sections with the involvement of a new secondary instructor. Violet was primarily involved in the veteran instructor's section of the course however she regularly interacted with both instructors. Structurally, the course ran as a typical introductory physics course at the university level with three 50-minute lectures per week, accompanied by one 2-hour lab section and one 50-minute discussion/recitation section. In contrast to a typical introductory physics course, the lectures and discussion sections involved extensive group problem-solving tasks that were designed to build connections between chemistry, biology, and physics (see [1,3,4,30] for more information).

The instructor in the lecture portion of the NEXUS/Physics class took a pedagogical approach that focused on attending to the disciplinary substance in students' reasoning [31] and supported students in formative scientific pursuits [32]. In practice, every meeting of NEXUS/Physics lecture offered multiple opportunities for student-centered group activities such as open-ended group white-boarding tasks, multiple-choice conceptual clicker questions with peer discussion, and weekly quizzes. The instructor invited alternative explanations and encouraged the discussion of different disciplinary goals (see [33] for an example). The discussion/recitation sections involved extended group problem-solving tasks that encouraged students to explore the connections between chemistry, biology, and physics. These problem-solving sessions were often designed not to lead students down a carefully pre-determined path, but as starting places for students to explore the realm of applicability of physics to biology and chemistry (see [2,3] for examples).

Violet at the time of this study was a junior enrolled in the biology major at the University of Maryland. Violet had declared a general biology major and was on the pre-medical school track. She was seen in her department as a strong student, hired to work as peer educator in introductory biology courses. She enrolled in the NEXUS/Physics Introductory Physics for Biologists course as part of the requirements of her major program. Violet had no physics class experiences prior to entering college. As mentioned in the introduction, we began following Violet as a way to explore developing competence in physics. However, upon initial analysis of the data we found that Violet's relationship with physics was more nuanced, and her relationship with physics at the end of the course felt starkly different than at the beginning of the course. Thus, we turn to analyzing the data with three research questions in mind: (1) How is Violet's incoming relationship with physics different from her outgoing relationship? (2) How can we explain the change between Violet's incoming and outgoing relationships? (3) How can the constructs of identity, affect, and epistemology provide access points for influencing relationships with physics?

### C. Situating data through ethnographic methods

We draw on a wide range of literature to characterize Violet's expanding relationship with physics. We use ethnographic methods to explore Violet's lived experiences with physics through analysis of classroom observations, artifacts from the class, and periodic interviews [34-36]. In the ethnographic work of this study, we gained direct access to two of three of the major components of the course Violet was enrolled in – the lecture and the recitation/lab. The authors were embedded in the lecture portion of the course as participant observers for approximately 35 hours taking field notes [37]. Violet's small group in the lecture and recitation/lab were also video-recorded for approximately 84 hours throughout the course. In the third component of the course (the course help center) we were not able to collect data, but through interview reflections we see it was a central feature of Violet's experience in the physics course. We paired these observer data streams with three one-on-one interviews with Violet, conducted with the first author. Figure 2 depicts our data collection throughout this study. This representation shows that we had the highest intensity of data collection in the classroom observations and video. These two data streams were often coupled to interpret the field note and interview data.

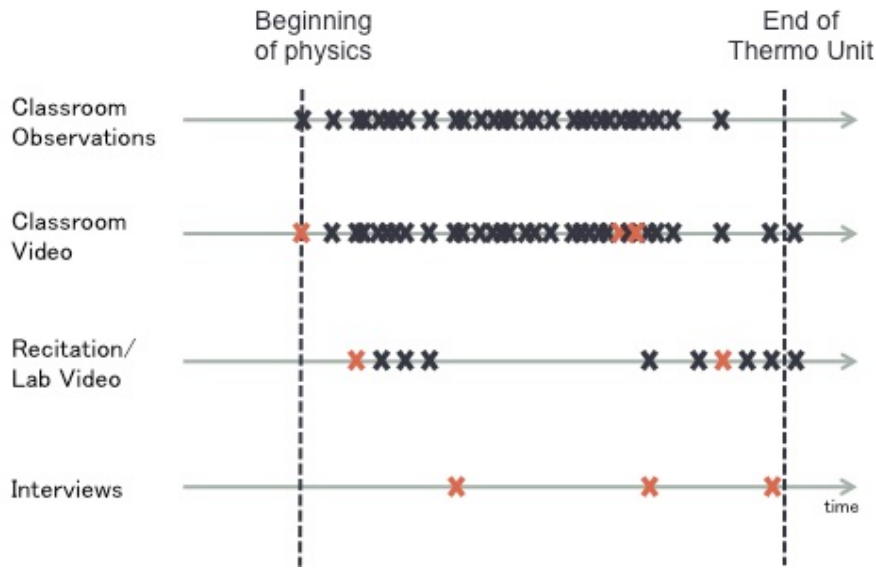

FIG. 2. (color) This figure is a schematic of the various data streams we collected. Time is represented along the horizontal axis and increases toward the right. The X's on the timeline indicate days we collected data in that particular stream. Orange X's represent data clips that are specifically analyzed in this paper.

# V. UNDERSTANDING VIOLET'S INCOMING AND OUTGOING RELATIONSHIPS WITH PHYSICS

Ultimately, our goal is to provide an account of Violet's developing relationship with physics. In order to meet this goal, we begin by independently examining her incoming and outgoing relationships with physics. We draw upon the theoretical connections between identity, epistemology, and affect to carefully describe the incoming and outgoing disciplinary relationships.

## A. Contrasting Violet's incoming and outgoing relationships

At a coarse grain size we argue that Violet's incoming stance is one of expecting to hate physics, and for physics to not be particularly useful to her. We see this stance evidenced in the finer-grained elements of how Violet positions her relationship with physics in classroom and interview moments in early parts of the class. On the first day of class, Violet publicly voiced her expectation that physics would "…make my life a living hell...as a biologist I've avoided physics for the longest time…to be a doctor you do need to know the physics aspects of the body…but I think it's just, you know, pain (Clip 1)." In an interview she reiterated this perspective saying,

> *(Clip 2) I feel like I came in with a lot of hatred towards physics in general....I never took [physics] while I was in high school, so coming in and knowing for the [biological sciences] curriculum that I had to take two semesters of physics, something that I never had any exposure to, I wasn't very happy about. And I'm more of a bio and chemistry person, and I just take physics for granted. I was like, oh it falls. OK sure. I think biology comes a lot more naturally to me than physics does. Because I think I can visualize the system better, having taken a lot more biology classes and chemistry classes. And I've just been exposed to more of it."*

From these starting stances it might seem to some that Violet is a lost cause. She expects physics to be difficult, describes it with hatred, and positions it in contrast to herself as a "bio and chemistry person." However, in observing Violet in classroom situations and following her in interviews we see that Violet's stance was in part a product of what she thought physics was about. ("Oh it falls. Ok sure")

The NEXUS/Physics course provided opportunities to dynamically interact with this stance, and ultimately changed Violet's relationship with physics. Throughout the NEXUS/Physics course, we created opportunities for physics to interact with students' biology knowledge. For example, Violet described one group problem-solving activity that asked students to consider the impact of modeling a DNA molecule as a spring. Reflecting on this task in an interview, Violet recounts how she helped explain the structure of DNA to the teaching assistant (TA).

> *(Clip 10 excerpt) And I went up to the chalkboard and like started drawing a string of DNA to show [the TA] because she was like I don't know any biology. And I was like ok I'll help you with this part, and I mean it just clicked so much more easier for me. I was like well DNA is blah, blah, blah there's this, this, and this. Maybe this factors into this and it takes more force for this to happen.*

In this quote we see how Violet's identification with and sense of competence in biology became an important part of how she interacted with this physics class. During the class, we saw multiple episodes of Violet bringing biology into her physics reasoning, with increasing confidence. By the end of the course Violet describes having a view about physics that was more connected with her knowledge from chemistry and biology,

> *(Clip 6 excerpt) Yeah, I feel like if I were in a regular [physics] class I wouldn't like the physics because it just doesn't relate. Because me as a biologist, I like biology and chemistry so much more than I did physics. But now that I can see the relationship between all three, it's kind of made me like physics more.*

At a coarse grain size, we see Violet's incoming relationship with physics as in tension with her relationship with biology. By the end of the course, however, Violet's relationship with physics has expanded to the point it beings to overlap with biology. This is not to say that the tension between the domains has been completely eliminated, but there is a connection between these two disciplinary domains. In the remaining sections we will elaborate on these contrasting incoming/outgoing relationships and explore the means by which Violet's relationship expands.

B. Violet's incoming relationship

In Violet's opening interaction with the course (as seen in Clip 1), we see that to some degree it is socially acceptable to for Violet to position herself as expecting physics to be "painful" (i.e. we would not expect to see students publically taking a similar position regarding reading, for example). This may serve to show that some of the initial stereotypes discussed at the beginning of this paper are not just voiced by faculty, but may represent some socially sanctioned ways of being a biology major. Several times in later interviews and in class observations, Violet expresses that she expected to hate physics upon entering the class and entered with an identification as "more of a bio person."

Violet also repeats characterizations from Clip 2 of her early views on physics as something that can be taken for granted, and does not require a more detailed explanation or justification: "I just get in the car and step on the accelerator and I go. I've never really thought about it, like oh the tires are exerting a force on the road and the road is exerting a force back," and, "Like the [direction of the force on a] pizza box [on the dashboard of a car], I was like well why don't we just get in the car and do it? Instead of writing about it?" This epistemological stance reflects a sense from Violet that physics considers primarily macroscopic phenomena that do not require explanation. This stance towards what counts as physics is repeated in early course interactions as Violet describes physics considering roller coasters going down really fast, dropping eggs, and rolling balls down hills.

In addition to these repeated characterizations of what counts as physics we see Violet exhibiting strong affective stances where she expects physics to be difficult and painful - to make her "life a living hell." In the opening day of class (Clip 1), Violet positions this stance in contrast to students who have suggested just before her that it might be relevant to biology to understand physics. This suggests that for Violet knowing in biology and knowing in physics are in tension with another. She also positions biology in contrast to physics when she says "as a biologist I've avoided physics." In the interview clip we see this affective stance (" I came in with a lot of hatred toward physics in general") repeated, as it is often throughout early segments of the class both in the interview and in in-situ data ("Oh my god, I hate this lab," and "I feel so stupid in this class"). These statements are commonly followed (as in Clip 2) by Violet positioning herself as "more a bio person."

Figure 3 demonstrates the ways we see Violet's relationship being modeled by the interaction between affect, identity, and epistemology in these early moments of the course. Clip 1 and 2 demonstrate how the affective stance of hatred toward physics is linked with Violet's positioning of herself as more of a bio and chemistry person in contrast to physics. This contrast between biology and physics is also reinforced by the stance that physics is primarily concerned with macroscopic phenomena that do not require further explanation. The consistent positioning of physics identity, affect, and epistemological stances in contrast to biology leads us to attribute some amount of "stability" to this incoming relationship with physics as strongly opposed to a relationship with biology.

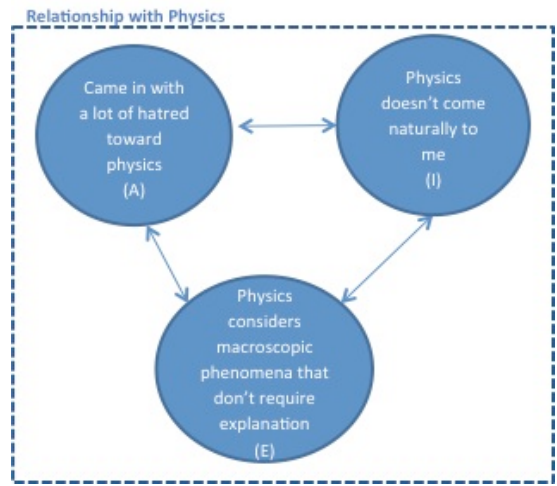

FIG. 3. (color) A representation of Violet's early relationship with physics as demonstrated by clips 1 and 2.

We also have some reason to believe that early portions of the class, before well-developed links to biology appear, reinforced incoming expectations for Violet. In the first recitation section for the course (a scaling problem on diffusion through an earthworm's skin; see Gouvea et al., 2013 for the full problem), the TA let us know that Violet was ready to drop the course which the TA interpreted as Violet being insecure about doing the math. In Interview 1 we asked Violet if she recalled this problem, and Violet replied,

> *(Clip 3)* *"Oh yeah, that was the first recitation...I remember I got really frustrated with this, maybe because it was like the second week of class and we only had a couple lectures. And when I'm seeing letters and equations usually it's a variable or a constant and it doesn't like stand for anything else. And then when they say like oh put this in like symbols, I was like [gasp] I don't know what to do. Oh my gosh like what's going to happen to me? I don't understand this. But I think afterwards it just took awhile for it to click like oh it's just a substitution problem, you're just substituting the letter in for what it stands for."*

Violet's recounting of this episode suggests that this experience reinforced her expectation to not be good at physics. She describes reflecting on this early situation as, "Oh my gosh what's going to happen to me?" By the time this interview happens we see Violet beginning to repair from that situation, "it just took a while for it to click." Nevertheless, the manipulation of symbols as an important part of participating in physics is reinforced here, and the experience did not improve Violet's sense of competence with the math. Throughout the course, Violet's ability to manipulate symbols becomes a point of focus for understanding her sense of capability in participating in physics.

In looking at moments where Violet was participating in class activities, we notice times where the group she is working with is doing productive sense-making around standard physics tasks, but Violet experiences a strong affective response (often negative). For example, in a lab section in the first month of the course, the students are working to construct kinematic graphs to describe an amoeba's movement over 3-minute intervals. Violet is working with a partner, Otto, to construct the graphs, and together they have recorded values for the position of the amoeba at different moments in time. At this point the two are working to figure out how to get the velocity graph from this information. Violet suggests that the slope of the position graph is the velocity, which she gets from a best-fit line, and calls the TA over to check her understanding.

**(Clip 4)**
*[1] Violet: (to Otto) So velocity is change in position,*
*[2] Otto: over time,*
*[3] Violet: over time.*
*[pause]*
*[4] Violet: So can we say that this [pointing to the computer] is the velocity? Of the graph? Of like the amoeba?*
*[5] Otto: So...sorry, what did you just ask?*
*[6] Violet: If this is the slope,*
*[7] Otto: which means the velocity.*
*[8] Violet: Yeah, that's what I'm saying. I'm going to ask [the TA] if that's the velocity.*
*[9] Otto: It is. Yeah, we'll have to...*
*[10] Violet: And then put the "t" as an "x" [in the equation]. The velocity would be blah blah blah times "x".*
*[11] Otto: The thing is that if you do a velocity versus time graph, it should look different, because velocity's not always the same. You know what I mean? We're not sure, I mean it's obviously not going constantly, because if it was it would all be going the same distance, right?*
*[12][The TA approaches the table where Violet and Otto are working.] Violet: (to the TA) Can we ask you this?*
*[13] TA: Sure.*
*[14] Violet: OK, this line - is that the velocity?*
*[15] TA: Ummmm, what velocity?*
*[16] Violet: The velocity of the amoeba.*
*[17] TA: Well, I know, I know. But which one, I mean the instantaneous velocity or the average velocity?*
*[18] Violet: Well, I don't know. Remember I'm physics stupid.*
*[19] TA: I understand. But what did you do here? What is that line?*
*[20] Violet: The slope of the graph.*
*[21] TA: Is it?*
*[22] Otto: Of the, of the position.*
*[22] [V sits back in her chair, orients her body away from the TA and looks down at her paper while Otto continues to talk.]*
*[23] Otto: So I mean, we understand it's the velocity of the position, and so what you want us to do is...So, should we look at each point and say OK what's the velocity from here to here? I mean we know velocity is changing with position over time. Should we do that individually?*
*[24] TA: So yeah, what you're asking is the difference between instantaneous and average velocity.*
*[25] Otto: Yeah.*
*[26] TA: What you have here isn't really either actually. This best fit line is kind of the average velocity. At least what you told it to do was to draw a straight line that was the closest to all the points as it can be.*
*[27] Violet: Yeah*
*[28] TA: There's a difference between that and the average velocity. The average velocity is the uh,*

[29] *Violet: Wouldn't it be [pointing to the graph on the screen] this minus this, divided by this?*
[30] *TA: Well, yes. That's more the instantaneous velocity.*

In this clip we see how Violet's incoming relationship with physics may have been reinforced by particular experiences early in the course. This episode opens with Violet and Otto doing productive work to think through how you find the velocity from a position versus time graph. In the first 10 turns Violet and Otto suggest that the slope of the graph should give them the velocity, and Violet suggests you could get that value from the best fit line. In turn 11 Otto expresses a concern that using this method would give a constant velocity, which prompts the students to call the TA over. The TA does not orient to the kind of reasoning that Violet and Otto had been doing, but asks them to distinguish between instantaneous and average velocity.

Figure 4 displays the cascading of events as they unfold in this clip. Consistent with her early descriptions from interviews, Violet positions herself to the TA as "physics stupid" (turn 18). Violet makes this positioning move immediately following the call for definitional knowledge (between instantaneous and average velocity) from the TA. As the TA continues to push for definitions of the velocity, Violet disengages from the conversation (turn 22). The sequencing of these events suggests that Violet's moves to distance herself from the physics (calling herself physics stupid and disengaging from the conversation) is in part a response from the TA's call for a particular kind of knowledge (definitional) as relevant and important to making sense of the situation. If many of these types of events occurred throughout the early parts of the physics course, we can see how the these early interactions may have drawn out particular ways of Violet positioning herself in relation to physics as well as reinforced her expectations about her relationship with the course.

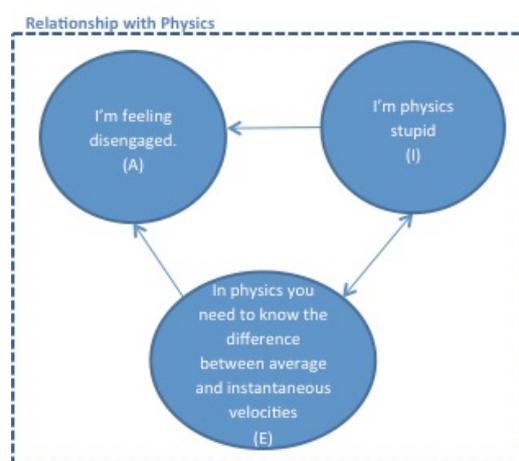

FIG. 4. (color) A representation of the cascade of events that unfold in Clip 4 in terms of affect, identity, and epistemology.

Across these various episodes, we characterize Violet's incoming relationship with physics as in contrast with her relationship with biology. (For example, in Clips 1 and 2 Violet notes that her avoidance of physics has been in part due to her being "a biology person.") In Figure 5, we represent these relationships by boxes with distinct boundaries. We note that in most cases Violet's relationship with physics arises in conversation as in tension with her relationship with biology (e.g. "as a biologist I've avoided physics for the longest time"). We represent this tension between the relationships with a jagged arrow between the two boxes. Within the physics relationship we note that the affect, epistemology, and identification with physics are all coupled (as we demonstrated in Figures 3 & 4). To describe how Violet sees physics as considering macroscopic phenomena that do not require explanation, while ignoring her disinterest in these knowledge elements would be a misrepresentation of her relationship. Similarly, the strong positioning of herself as not identifying with physics is coupled with the stance on what counts as knowledge in physics, and her emotions tied to that stance.

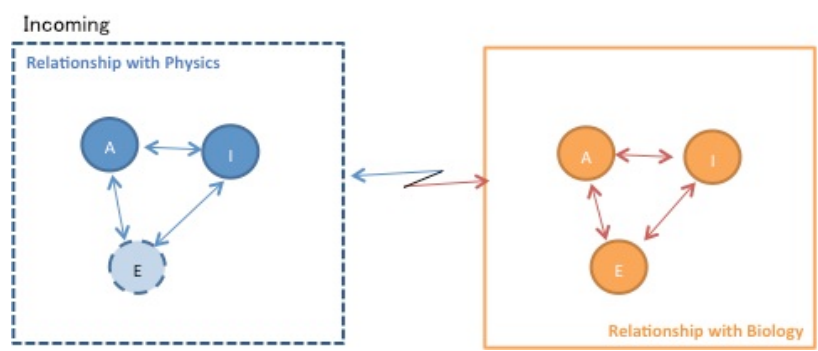

FIG. 5. (color) Violet's incoming relationship with physics is in contrast with her relationship with biology. The jagged arrows represent tensions between the constructs.

### C. Violet's gradually expanding relationship

As Violet continued in the NEXUS/Physics course, we continued to follow her progress, timing 2 additional interviews to take place at the end of the first semester of the course - at the end of the chemical energy unit (see [1] for more information), and again at the end of the unit on entropy and the second law of thermodynamics. These later data streams provide additional evidence for the stability of this incoming relationship with physics. They also demonstrate an expanding relationship with physics reflected in Violet's affective stance toward physics and in her ways of engaging in symbols work and graphical reasoning.

In the lecture portions of the class, we see glimpses of Violet's developing sense of competence with physics, and a productive level of participation. For instance, on two different days during the last portion of the first semester course we see Violet express confidence in answers to clicker questions. One of these comes from a day when the students have been studying fluid flow, and the question is about the main blood vessel in the heart being split into arteries. When the question is posed we see Violet shout out to her group, "Oh! I know how to do this!" and she proceeds to the take the lead in working out the problem with her group. In doing so she explains that the "n" in the algebraic expression that she has written (nAv = nAv) represents the "number of arteries. Aorta has one, and it's splitting into two," as she replaces the numbers in the expression.

In the second example, the question is about the force felt by atoms in a bond at a particular distance (2 nm). In this case the majority of the class has indicated that the correct answer to the question is 0 force. The professor emphasizes this, and Violet interjects emphasizing her reasoning in a matter of fact tone,

**(Clip 5)**

*Prof: Alright, most people are saying zero.*
*Violet: The answer* is *zero. {laughter}*
*Prof: You know what the answer is?*
*Violet: Yes.*
*Prof: How come?*
*Violet: Because if you go to 2 (the distance referred to in the question) and you go all the way down. Ok. It's the slope of the graph at that point is zero. Which means the force is zero.*
*Prof: Beautiful.*
*Violet: And that means the atoms are in equilibrium with each other.*

These examples show Violet readily engaging with the symbols work and graphical representations, and making meaning of the representations in terms of the physical picture. These snippets demonstrate the sense of competence that Violet is developing, as well as how her willingness to engage with symbolic reasoning has developed in this course. In both of these clips she expresses a great deal of confidence in her reasoning ("I know how do this!" and "the answer *is* 0."), while also demonstrating that she understands that what counts as sufficient reasoning is not just the numbers, but the reasoning behind them. When we look at Interview 2 data, we see corroboration with the ways she positions herself in relation to physics.

*(Clip 6)*
*[1] Interviewer: So tell me just a little bit about how, what's your impression of the course? That you've taken, you're almost done (the first semester is almost completed).*
*[2] Violet: Um yeah, well I've definitely changed my answers on that survey [the Maryland Expectations Survey [38,39]] ...I was like I think I like this course*

> *better now because I actually see the implementations in the biology rather than like no biology at all. Cause I like these questions like the ATP question and the heart blood rate flow. I was like ok that's cool it actually relates.*
> *[3] Interviewer: So you feel like when you came in you didn't expect it to do that?*
> *[4] Violet: I didn't expect it to, I don't know, I don't know. I just like it more because I can see the relationship now. I think like at first when we were talking about like forces I didn't really see like -oh the forces on the body, but then we did flow and now we're doing energy. It's like ok that relates a lot better than the other stuff that we were talking about. So I like it.*
> *…*
> *[5] Violet: Yeah, I feel like if I were in a regular [physics] class I wouldn't like the physics because it just doesn't relate. Because me as a biologist, I like biology and chemistry like so much more than I did physics. But now that I can see the relationship between all three, it's kind of made me like physics more.*

In clip 6 we see corroboration with the ways Violet participated with the physics in other settings. As she describes in turn 4 she see ideas of energy and flow relate more, particularly in the application to biology (turn 2). Similar to earlier points in the data, we see Violet strongly positioning herself as a biologist (turn 5), but in contrast to earlier points we do not see her positioning that in contrast to physics. Instead, here she draws attention to the connection between all three disciplines. She claims that seeing connections between the disciplines impacted her affective stance toward physics such that "it's kind of made me like physics more." This suggests that as the tension between biology and physics decreases, other components of the relationship are changing as well.

## D. Violet's outgoing relationship

As Violet moves through the thermodynamics unit of the 2nd semester of the course, we continue to see her relationship with physics expanding to include new elements. However, we hesitate to characterize this change as a distinct shift. It is not so much that Violet leaves behind her old ways of describing and relating to physics, but rather that her relationship with physics continues to evolve and expand.

In Interview 3 with Violet two-thirds of the way through her year-long physics class (see Figure 2), Violet recounts a moment from class when she felt successful. Violet describes a class period where she asked the professor if you could manipulate the Gibbs free energy expression ($\Delta G = \Delta H - T\Delta S$) in a particular way, and the professor affirms her proposed manipulation. Violet finishes the recounting of this experience in the interview by saying, "It makes me feel so glad that [I] get what's going on." The interviewer asks her to elaborate on this moment from class. Violet then describes working with the Gibbs free energy equation, closing with, "We were just figuring out the different equations for enthalpy and entropy and what they rely on, so it's like delta G is basically equal to the change in internal energy plus work minus heat." The interviewer asks if this was helpful to Violet and she replies,

> *(Clip 7)*
> *[1] Violet: It is [useful] to see that overall. Because the delta-G that you end up with is what people use to change the internal energy and do work on the system.*
> *[2] Interviewer: So this just helped you see how that was happening?*
> *[3] Violet: Yeah, conceptualize it all in symbols. Cause I remember last semester I was not a symbols person. And even today, when we were doing recitation I was like, "Why are there so many? Why is there one symbol for so many things? It's so confusing. U is internal energy and potential energy, and UGHH I hate this." It's really frustrating.*
> *[4] Interviewer: But you just said that* last *semester you weren't a symbols person.*
> *[5] Violet: Yeah, I'm getting more used to it.*
> *[6] Interviewer: You're getting more used to it?*
> *[7] Violet: As the time goes on and you get more symbols.*
> *[8] Interviewer: So was this [moment from class] significant to you because it was reasoning about the symbols?*
> *[9] Violet: Yeah, and it helped like put the word with the symbols after like working with them for so long. It's like you see the equation, but what does the equation actually mean?*

Notice in this short excerpt Violet's own awareness of her shifting relationship with physics. In turn 3 Violet describes how *last semester* she was not a symbols person, suggesting that she sees a shift in the ways she identifies with symbols work. She goes on to describe a situation where she manipulated symbols effectively in the equation, and then proceeded to make sense of "what does the equation actually mean?" In the epistemological stance Violet adopts in this moment, we see glimmers of productive views on knowing and learning [40] coming to bear; Violet tags understanding what the Gibbs free energy equation means as interpreting conceptual ideas including "work" and "internal energy." She goes on to demonstrate a sophisticated understanding of what the equation means:

> *[10] Violet: You get this equation in chemistry and biology and now in this class. But you learn it from all different ways, all different angles. And I feel like in this class it's so much combined with biology, you have to put those two realms together, and sometimes it could be totally opposite but sometimes it comes together.*
> *[11] Interviewer: So was this an example of opposite or coming together?*
> *[12] Violet: Coming together definitely, cause in bio you're always like oh delta-G is how much free energy you have to do work in the system. And now in this class you actually have a specific definition of what work actually* is*, and instead of just oh it can make this product, but you can see how an enzyme fits into what work is. I mean we haven't touched on that but it's like you're changing configurations and, which needs a force, so it coincides with it.*
> *[13] Interviewer: And so enzymes are connected with this equation [delta-G] in biology?*
> *[14] Violet: Yeah I mean like overall, an enzyme is used to catalyze a specific reaction, and that specific reaction will have a delta-G that corresponds with it. ... So we'll use the catabolic processes and the energy that it releases to be able to do work to build up the bigger macromolecules.*
> *[16] Interviewer: But now you feel like you have a more specific understanding of what that work is?*
> *[17] Violet: Exactly, because when the enzymes come together, and they bring the products and the substrates together they're interacting as a force that will shove* *the two things together, so the reaction does occur more fast. And the enzyme, in order to become active, needs to usually change its conformation, which can be equivalent to changing a displacement.*

We see in turn 12 that Violet sees making sense of the symbols in physics as important for making ideas from biology, chemistry, and physics come together (i.e. Violet values the particular epistemological stance of the knowledge from the disciplines being connected). Making sense of the equation is tied to both unpacking the conceptual ideas the symbols represent, "the delta-G that you end up with is what people use to change the internal energy and do work on the system," and the additional understanding gained from these ideas, "now in this class you actually have a specific definition of what work actually *is*," as "a force that will shove the two things together, so the reaction does occur more fast." There is still some evidence that Violet struggles to keep track of what meaning she should associate with a symbol in a particular moment (turn 3), but there's also a sense that she is trying to extract meaning from the symbols about what work really is and what it means for an enzyme in a chemical reaction. We point to this clip as evidence that Violet's epistemological stance towards physics knowledge is expanding. We see this stance as expanding in two ways: 1) in the way that she pursues the physical meaning of symbols and 2) in the way she sees physics knowledge as consequential for understanding things in chemistry and biology. We have chosen to represent this second way as "outside" her relationship with physics because, although this stance is linked to knowing in physics, it is simultaneously linked to ways of knowing in biology and chemistry (in this way it does not live solely within her relationship with physics). Figure 6 depicts these connections between affect, identity, and epistemology reflected in this moment.

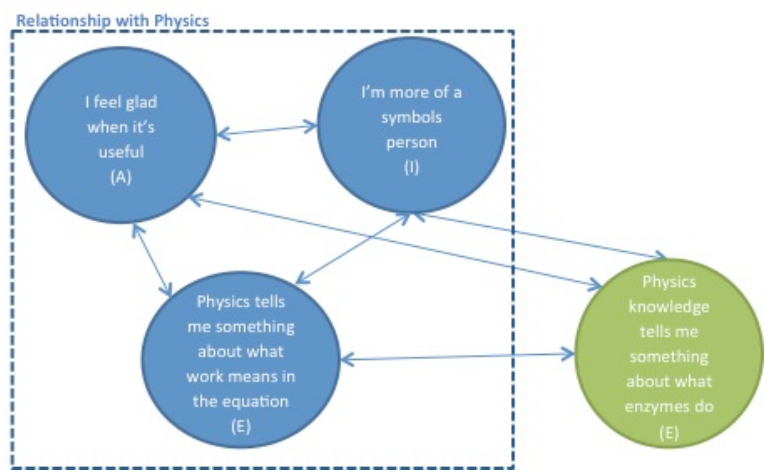

FIG. 6. (color) A representation of the scientific reasoning moment Violet describes in Clip 7. The green circle outside of the relationship with physics indicates a developing epistemological stance that physics is relevant to biology knowledge.

Across this entire clip we also notice how these ideas of being a symbols person, making sense of equations, and connecting physics knowledge to biology are tied with the affective experience Violet is recounting. She begins the description of this event by saying, "It makes me feel so glad that [I] get what's going on." In the moment of her describing this to the interviewer she also relates a sense of pride in explaining how it all connects. She says to the interviewer just after turn 17, "I love all this biology." We see ties between Violet's affective stances of feeling glad and loving the biology, the epistemological stance about connecting ideas across the disciplines, and her identity as a biology person.

Clip 7 suggests that Violet's outgoing affective stance toward physics is more positive, and provides additional evidence of Violet's incoming stance of "hatred" towards physics. Her outgoing relationship, while different, is not one of loving physics, but rather more of a love-hate relationship with physics:

> ***(Clip 8)*** *"I don't know, I mean, before I took physics, I guess I usually hated it because I never took it before so I didn't know what to expect. I mean some days I really do still hate it, because I just get really frustrated, but I think it's more of a love-hate relationship than just a pure hatred relationship."*

Later the interviewer probes moments when Violet still feels frustrated with physics,

***(Clip 9)***
*Violet: Physics isn't something that I can pick up and can understand really easily, and like he'll [the professor] go through the PowerPoint, or he'll go through the clicker question....But it's like, physics for me, you need to go over it over and over, constantly just to make sure that you have the equations down right, cause I think in biology it's such specific methods that you use all the time. You learn a pathway, you learn a reaction…a lot of the reactions in your body work using the same mechanism. But from the physics side, you can use one equation, to explain 50 different things, and each time the equation will mean something different.*
*Interviewer: Can you give me an example of an equation like that?*
*Violet:....like in the recitation, it was like, "How does the energy U change?" And I was like, "Is this U internal energy? Is this potential energy?" There's so many things used to describe one symbol, and that's where it can get confusing, if you're not really specific about what you're talking about. And there's a lot of different ways to derive the equations… And I feel like in biology it's like one way to come to the conclusion, one way. It's only this pathway that can happen, it's only this reaction that can take place in chemistry, whereas in physics you can come at it from so many angles.*

From these two clips (8 and 9) we can see some of what frustrates Violet with physics. She describes physics as something that does not come easily to her, and requires her to spend time going over it again and again. The frustration Violet describes is connected with her views on knowing and learning in physics. She describes how "you can use one equation to explain 50 different things, and each time the equation will mean something different" indicating that she sometimes sees equations in physics as having many different meanings that you have to be able to interpret. This epistemological stance is one that [40] would associate with a fragmented view of physics. However, we also see in clip 9 Violet contrasting physics with biology and chemistry, in which explaining a phenomenon is straightforward with one solution, while in physics there are many different derivations and starting places. This epistemological stance of seeing multiple pathways to solutions in combination with Violet's symbols work in clip 8 (describing the importance of understanding what an equation really means) suggests that Violet is also developing a set of more sophisticated views of physics [40]. We see these different epistemological stances as indicative of a mixed relationship with physics. Frustration is tied to a fragmented view of physics knowledge and a view that physics has many different pathways to solutions. These views are also connected to a sense that physics isn't something that comes easily to Violet. This description is reminiscent of the incoming relationship with physics that is in tension with Violet's relationship with biology. However, we also see an expanded set of coherences as depicted in Figure 6 that suggest the connection between Violet's relationships with physics and biology are less in tension with each other than in the earlier moments.

Violet's outgoing relationship with physics is clearly different than her incoming relationship with physics. In the outgoing relationship we see her affective stance towards physics moving between frustration with physics and an appreciation of the usefulness of physics. This mixed stance is consistent with classroom observations of Violet moving between seemingly enjoying physics problems and also making occasional meta comments about hating physics. Tied with this mixed affective stance is how Violet positions herself as more a symbols person, but simultaneously feels that physics is easier for other people than it is for her. This suggests that her mixed affective stance may be associated with shifting how she is identifying with physics. Her epistemological stances toward physics, tied sometimes to appreciation and sometimes to frustration, have also become more varied through seeing equations as meaningful and useful, while still seeing a variety of meanings in those symbols.

VI. BY WHAT MEANS DOES VIOLET'S RELATIONSHIP WITH PHYSICS CHANGE?

Up to this point we have described the ways that Violet's incoming and outgoing relationships with physics are distinct. Now we would like to turn to asking how the shift in relationship came to be. Going into the analysis of how Violet's relationship with physics changed, we expected to build off earlier work [41] and attend to the opportunities that Violet had to develop her sense of competence in physics. This literature suggests that a way to mediate students' expectations about being incapable in physics is to provide them with opportunities to see themselves as successful, to see others being successful, and to receive feedback on their successes. We do see elements of this story in Violet's changing relationship with physics – as she describes having successes in physics and receiving feedback about her achievements in the course. However characterizing these experiences solely in terms of opportunities to develop competence in physics misses a key component to the story of Violet – her relationship with biology.

### A. Violet's Positive Affect and Strong Identification with Biology is a Lever for Influencing Her Relationship with Physics

In the early interview, Violet does not yet have a clear articulation of what it means to know or learn in physics other than simply seeing physics as something that you can "take for granted" and that doesn't require explanation (clip 1). On the other hand we see Violet describing that biology "comes a lot more naturally to me than physics does," and identifying as "more of a bio and chemistry person". This feeling of a strong tie with biology becomes consequential where we see Violet engaging more with activities that have an obvious biology content connection (e.g. examining DNA as a spring, or connecting ATP and energy) than with activities that solely reside in canonical physics (e.g. pushing a box across the floor, or pressure in a tank of water). In interview reflections these moments become salient as Violet describes feeling competent. For example, in the interview setting Violet reflects on the "modeling DNA as a spring" task that she had completed the same day in the recitation. Violet begins recounting how she helped explain the structure of DNA to the TA as an example of how she is knowledgeable in biology while others know more about physics,

> *(Clip 10)*
> *[1] Violet: And I went up to the chalkboard and like started drawing a string of DNA to show [the TA] because she was like I don't know any biology. And I was like ok I'll help you with this part, and I mean it just clicked so much more easier for me. I was like well DNA is blah, blah, blah there's this, this, and this. Maybe this factors into this and it takes more force for this to happen. So it was easier for me to explain that side and for her and other people in the class the physics just comes so much more naturally.*
> *[2] I: Mhm, but just now you were saying so maybe when the force does this. So do you think that your feeling like you're good at doing the biology helps you do the physics part at all?*
> *[3] V: I think it did today.*
> *[4][Interviewer pulls out a copy of the recitation sheet for DNA as a spring Violet is referring to] I: This one, right?*
> *[5] V: Yeah and the questions were, we were looking at this graph and we were trying to say oh what does the DNA look like at certain regions on this graph. And then I was explaining to people, in the cell DNA looks like this at one time, but then at a certain time it can look like this. So if you're trying to stretch the DNA it probably starts out in a more condensed form and then expands. Rather than going from a gobbely goop of spaghetti down to something really tightly compact which normally happens. And so I was drawing out on the piece of paper it looks like it goes from sister chromatids to something that's less condensed and less condensed and less condensed, and eventually at a certain point the hydrogen* *bonds are going to split and then the backbone is going to split. And so that's where I think my biology came into play, like knowing the structure and then more on [the question from the worksheet], "What properties of the DNA might determine the force?"*

[6] *V: And I was like oh, different bases when they base pair to each other they have different number of hydrogen bonds, so the greater number of hydrogen bonds it has you know the more force you're going to need to break the bond, the more energy you're going have to put into breaking them. And I gave them an example of saying well like you know during DNA replication the helix opens at a point where the DNA is very A-T rich because there's less hydrogen bonds to break hence you don't have to use that much force to break them.*

[7] *I: So that's why you might need less force?*

[8] *V: Yeah. And then we were saying oh if the backbone breaks where exactly will it break? You know? Will it break where the oxygen is attached to the phosphorus or the oxygen attached to the carbon? What's the bond strength going to be?*

[9] *V: And then for the last question where it's like [reads from worksheet] "When is it biologically important that the DNA not be under tension?" We were saying oh during DNA condensation when you know genes don't have to be expressed because there are some genes that are not expressed in the cell at all, so that part of the DNA is going to be tightly condensed. And when is it important that it be stretched? Well, during replication, transcription, translation when you actually have to work with that specific gene. And then there's no tweezers in the cell so what actually you know causes the DNA to do this? Well it's the proteins. So the proteins act as the tweezers to pull the DNA apart.*

In this clip we see how Violet's relationship with biology became an important part of how she interacted with this physics class, as depicted in Figure 7. We have chosen to represent this epistemological stance of knowledge from biology as relevant for reasoning in physics as "outside" her relationship with biology because although this stance is linked to knowing in biology it is simultaneously linked to ways of knowing in physics (in this way it does not live solely within her relationship with biology). Violet opens this example in clip 10 by contrasting her competence in biology with her competence in physics, and positions herself in turn 1 as knowledgeable in biology in contrast to "other people" in the class who find physics comes more naturally. In this clip Violet describes contributing facts about biology (the structure of DNA in turn 5 and the process of replication in turns 6 and 9), and the role it plays in different moments to thinking about the questions about force from the recitation problem. This type of activity, where Violet contributes knowledge about and drawings of biological structures, is a regular occurrence in her participation in the physics class for group problems that have a biological relevance. In this clip we see evidence that the TA affirmed that contribution, as Violet described herself showing "[the TA] because she was like I don't know any biology." We suggest that Violet's positive affect toward and strong identification with biology in this episode is a productive lever for influencing how she engages with the physics task.

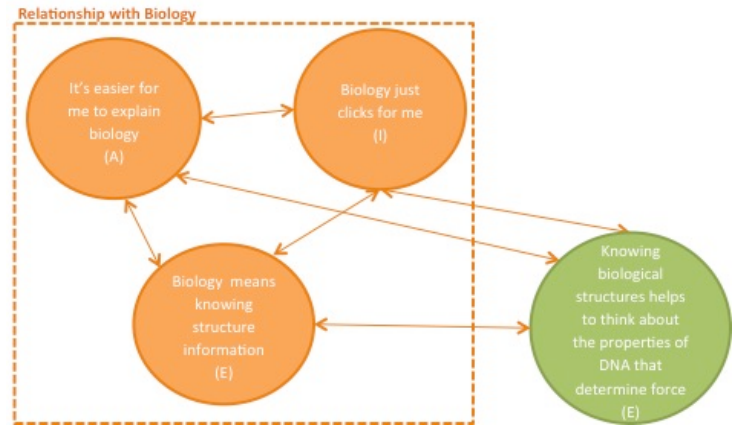

FIG. 7. (color)Violet's relationship with biology is depicted here in the ways that she describes it playing a role in her work in clip 7.

### B. Making Violet's Relationship Meaningful to Succeeding in Physics Impacts her Relationship with Physics

We present clip 10 and 11 as examples of the ways we see Violet participating in physics problem solving across particular kinds of tasks in the course. In tasks that were designed to foreground connections between physics and biology, we regularly see Violet identifying strongly with biology while contributing to the sense making around the physics. In these classroom episodes we most commonly see Violet responding positively to invitations from her group member to provide facts from biology to the problem solving (including structural diagrams, or process details). For example, in a recitation task focused on understanding how a kinesin motor protein uses ATP to move, we see Violet offering interpretations from biology to help make sense of the physics interpretation. In this example the students are asked to estimate how much energy is required to move a kinesin protein down the length of a leg if each step of the kinesin covers 8.5 nm.

*(Clip 11)*
[1] *Violet: I'm saying my leg is one meter, which is $10^9$ nm.*
[2] *Betsy: So when it steps it covers 8.5?*
[3] *Violet: Yeah.*
[4] *Betsy: Ok.*
[5] *Violet: So I divided that by 8.5 cause like it says one ATP moves it 8.5.*
[6] *Betsy: OK, I got it.*
[7] *Violet: (Doing the calculations on her paper) So that's ATP...*
[8] *Betsy: ...and that's kilojoules.*
[9] *Violet:...and then I multiply that ...*
[10] *Betsy: So, (reading the next prompt) finally discuss in your group what it means to say that a cell uses ATP to do molecular work.*
[11] *Violet: It uses the energy it gets from ATP hydrolysis...*
[12] *Betsy: So, the way we think of it is the motor head binds to it, and eventually it breaks the phosphate group off.*
[13] *Violet: I mean yeah. Like active transport think of the sodium potassium pump, right? The ATP will bind to the protein, and then once it's hydrolyzed it can change its shape, and then the ADP will be released, and then it will change its shape again.*

*[14] Betsy: Good. We understand.*
*[15] Violet: I mean we're biologists, guys. (pause) We are* biologists *(fist pump).*

In this clip we see Betsy and Violet working together to determine the amount of energy in kilojoules that the kinesin would use to move the length of the leg (Note: both the units and the modeling of energy in this task differ from many canonical biology tasks that may require students to do a currency analysis based on counting the ATP units "used" and the resultant energy "produced" during a hydrolysis reaction). In turn 10 when Betsy reads the physics task prompt from the worksheet about "what it means to say a cell uses ATP to do molecular work," Violet takes this up as invitation to contribute her biology knowledge in the context of this problem solving task in physics. In turn 13 she offers reasoning about the sodium potassium pump example as an analogy to the kinesin active transport example they are considering, which Betsy positions as evidence that they understand. Violet finishes this sequence in turn 15 by positioning herself (and her group mates) as biologists. Figure 8 demonstrates how knowing in biology links to a sense that biology knowledge is connected to and relevant for solving problems in physics. This positioning of relevant knowledge bits from biology in solving a physics problem supports the positive affect associated with being a knower in biology, and the positioning of the group as biologists. Episodes like this and Clip 10 recur in the physics class, giving Violet repeated opportunities to be positioned as a knower in biology, having a strong identity with biology, and enjoying contributing details from her biology knowledge to the problem solving in physics. We argue that Violet's relationship with physics expands as her relationship with biology becomes meaningful to success in physics.

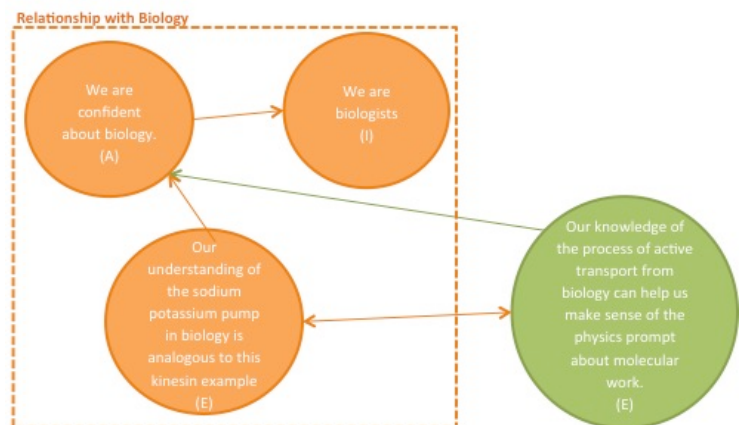

FIG. 8. (color) A representation of the cascading of events that occurs in clip 11 as it fits into understanding Violet's relationship with biology and the developing epistemological stance of biology being relevant to solving problems in physics.

## C. A Sense of Connectedness between the Domains of Physics and Biology Impacted Violet's Relationship with Physics

In Violet's outgoing orientation, the relationships with physics and biology are positioned in less oppositional ways, and a new epistemological stance develops around the expected connections between physics and biology knowledge. In Violet's outgoing stance, she sees physics as mattering for things that she cares about (i.e. physics contributed to a deeper understanding of what work is and ways of understanding what equations mean, which helped her make sense of enzymes). In the $\Delta G$ example (clip 7) we see Violet working to assess whether or not things from physics and biology are coinciding. For Violet, the example of working with the $\Delta G$ equation in physics is given as an example of things "coinciding" across physics and biology. These pieces of evidence suggest that her outgoing relationship with physics is no longer characterized in direct contrast to her relationship with biology, but rather she now sees the domains of physics and biology as linking up with each other.

We argue that the connection between the domains of physics and biology is the lever that ultimately influences Violet's expanding relationship with physics. Initially, Violet displays a bit of anxiety around the mathematical reasoning required in physics, expects physics to be painful, and thinks of physics as focused on things that just happen and do not require explanation. However, as she engages with a course that has been designed to highlight the connections between physics and biology we see Violet's relationship with physics begin to change. In understanding Violet's shifting relationship with physics, we see new elements developing – elements that represent an expectation that physics and biology knowledge should be relevant to each other. We posit that this expectation emerges from Violet's interaction with the NEXUS/Physics course that has been redesigned to highlight the connections between physics and biology. The ways that Violet describes using her biology expertise as a lever for understanding physics questions, and using physics to understand biological phenomena (the green bubbles in Figures 6, 7, and 8) provides evidence that in this environment, physics and biology are blending and the blend is consequential for Violet's relationship with physics.

As Figure 10 depicts, in the outgoing account we see an expansion of Violet's relationship with the domain of physics, to the point where it overlaps with the domain of biology (as depicted by the overlapping boxes in the outgoing relationship). There remains some tension in these relationships overlapping (e.g. Violet says she says "Seeing the relationship between [physics, chemistry, and biology] has kind of made me like physics more. Let's keep that secret between us."), but there is a salient connection between biology and physics knowledge. Drawing on this emerging connection we see how Violet's relationship with physics and relationship with biology begin to overlap. For example, in the $\Delta G$ clip (clip 7), we see Violet describe feeling "glad" at least in part about connecting her biology knowledge of enzymes to her physics understanding of work. In modeling DNA as a spring in a physics class (clip 10), Violet describes her expertise with DNA as contributing to success in modeling how the force required to change the configuration of DNA would vary based on the structural form of the DNA. These moments indicate that Violet's emerging relationship with physics has been influenced by the opportunities in physics class to productively draw on her biology knowledge.

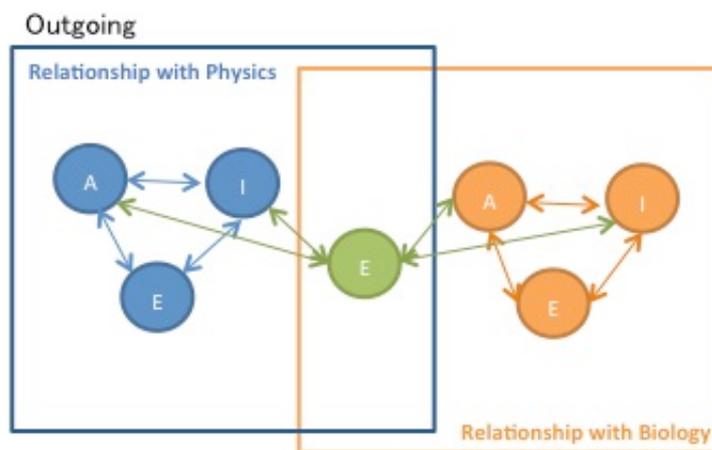

FIG. 9. (color) Violet's outgoing relationships with physics and with biology. In this representation the overlapped orange and blue boxes indicate the relationships beginning to overlap. This is connected with the green epistemology bubble in the overlap, indicating an epistemological stance that physics and biology knowledge are related. Within each relationship box the affect, identity, and epistemology bubbles are linked.

While we argue that Violet's relationship with biology was influential to her changed relationship with physics, we do not want to leave the reader with the sense that Violet only productively engaged with physics when the context was clearly biological. Rather, we are drawing on these overlapping domains to model how her relationship with physics in general changed. We point to examples of Violet reasoning in pure physics contexts (such as Clip 5) as examples of how she engages with physics-specific contexts. We believe that the repeated opportunities to connect biology knowledge with physics knowledge reinforced and sustained Violet's outgoing relationship, but we also believe her disciplinary relationship with physics was productively impacted. While we do think Violet could be tipped into a mode of being frustrated with physics and finding it to be concerned with things she doesn't care about, we also see cases of Violet productively engaging with physics that do not have a strong biology connection.

VII. INTERACTING WITH STARTING POINTS FOR LIFE SCIENCE STUDENTS

In understanding Violet's shifting relationship with physics, we see new elements developing – elements that represent an expectation that physics and biology knowledge should be relevant to each other. We posit that this expectation emerges from Violet's interaction with the NEXUS/Physics course that has been redesigned to highlight the connections between physics and biology. In the data we have presented here we are drawn to the ways that the NEXUS/Physics course supported creating a interdisciplinary space connecting the disciplines of physics and biology [3].

The relationship construct as we have proposed it is helpful here as we can see that there are multiple points of entry for creating these interdisciplinary spaces with students. Tapping into students' sense of self as biologists, for instance, may bring along with it a sense of competence and enjoyment in the process. With these two resources tapped, we may then be able to create situations where the discourses of physics and biology are brought into interaction and impact the affinity for physics. Our work suggests that in designing learning environments we should consider carefully the identities and affective orientations students bring into our classrooms, particularly at the advanced levels of students' academic trajectories. The case of Violet provides insight into how a student can feel expert in a classroom without feeling like an expert in the discipline.

We have a responsibility as curriculum developers and educators to meet students where they are and become partners with students in accomplishing transformations as part of the education process. It is difficult to meet this goal if we continually tell stories that assume students are not capable and have little hope of developing competence. We also will have difficulty meeting this goal if we put the sole burden for the transformation on the students themselves (e.g. students need to work harder, put in more time, arrive with more mathematical expertise, etc.). When we think of our students as a list of characteristics that focus on the pieces that they are missing we lose the opportunity to recognize the strengths they bring to our classrooms.

In analyzing the case of Violet and her developing relationship with physics we have demonstrated how opportunities to draw on her biology knowledge, affinity for biological systems, and sense of self as a biologist impact how she engages with and participates in physics. While we cannot do an in-depth case study of every student in a classroom to identify the strengths that each student brings, we believe that developing a diversity of such change trajectories would be a worthwhile future endeavor. We have reason to believe that some of the strengths we have identified with Violet are ones that other students in an Introductory Physics for the Life Sciences may share. Many of our students enter the classroom with strong biology backgrounds, affinities for biological phenomena, and identify as biologists. We suggest that an interdisciplinary space like the NEXUS/Physics environment that Violet interacted with would be a productive space for engaging many of our students. In this section we put forward some of the structural characteristics of the NEXUS/Physics course that we believe demonstrate what it looks like to become partners in the transformation with our students. Based on our efforts over three years to develop partnerships with students in this transformation, we have found guidelines that we find helpful in communicating with faculty about what they should attend to in creating similar environments. We present these as three design principles for creating spaces for interdisciplinary learning in which students can see opportunities for leveraging their strengths in making progress in physics.

A. Principle 1: Give students opportunities to share their complementary expertise

As we demonstrated in the Violet case, there were multiple opportunities for students in the NEXUS/Physics environment to share their expertise from biology. In the reasoning about DNA as a spring example (clip 10) Violet reflects on one of these moments where sharing information about the structure of DNA was consequential for modeling DNA as a spring. Opportunities like this were common in the NEXUS/Physics class, and students were encouraged to share their expertise - particularly in cases when the instructor(s) were not as knowledgeable as the students. We do not mean to suggest that instructors should give all responsibility for content knowledge over to the students, but we have seen that giving space to students to offer complementary expertise provides opportunities for students to make new connections and see coherence between different sets of knowledge.

This principle on the surface seems straightforward – build on students' existing knowledge. However, in the case of teaching physics to biology majors we have found that opening space for the students to be the experts on a topic is a powerful tool. In many instances throughout the NEXUS/Physics course we see the instructor inviting the students to contribute details or background knowledge from biology and chemistry (e.g. when the TA invites Violet to share information about DNA). The consequence of these interactions is that students begin to see past experiences in other disciplines as valuable and important for making progress in physics. This can be difficult for instructors because it means letting go of some of the control in the classroom, but it can also be freeing because it suggests that physics instructors need not become experts in biology before creating this type of classroom space.

### B. Principle 2: Make sure that driving questions are compelling to students

In connection with the previous principle, it is insufficient to give students the opportunity to share their expertise unless that expertise is consequential for making progress on a problem. In the case of Violet we see her orienting to a particular problems (e.g. the structure of DNA in thinking about forces and the work that an enzyme does) as compelling problems that require her biology knowledge. However, in these instances it is not only that these problems draw on her biology background, but that this background is consequential for making sense of the physics problem (e.g. the force required to stretch DNA, the meaning of work). From moments like these derives the principle of focusing the driving questions in class on problems students find compelling. We have identified many of these problems by listening to students and focusing on places they seek coherence (see [1] and [2] for examples). We contrast this perspective with one that focuses on identifying biological application problems or problems that superficially provide motivation using biological scenarios without using that knowledge to make progress on the physics to be understood [3].

### C. Principle 3: Give accolades to students' for successes and/or progress made

Another design principle we extract from this process is providing authentic feedback to students for the successes they demonstrate and progress they make. In the case of Violet we see her reference not only moments where her expertise was leveraged, but also describing praise from her instructor(s) as playing important roles in her evaluation of those experiences. In the example from class where Violet reasons about the force on the atoms in the potential well, the professor responds to Violet's reasoning with praise (clip 5). In the DNA example (clip 10) we see Violet reference the TA affirming her move to provide information about the structure of DNA, and she describes the affirmation from her professor about making sense of the symbols in the delta G equation (clip 7). While we do not attribute all of the progress that was made in Violet's relationship with physics to these expressions of praise, we do see them as positive indicators of Violet's progress. These accolades help students recognize when they are doing productive work that we value in the classroom, and may help them distinguish between bringing in valuable biology knowledge and bringing in bits of information that are tangential to making progress on the problem at hand. For example, in the NEXUS/Physics class we often heard the instructor saying things like, "Often when you are wrong, you are not completely wrong, so I don't want you to throw [away] your reasoning" in order to encourage students to see their ideas as valuable and worth building upon.

These three principles provide some guidance and support for creating spaces for interdisciplinary learning in which students can see opportunities for leveraging their strengths in making progress in physics. Yet it is important to remember that the goal in this endeavor is to become partners with our students in these transformations. The NEXUS/Physics course has started down the path of creating resources for instructors who are interested in building these courses, but there is still much to learn from this process. We call for future researchers to consider how their students are responding to attempts to make connections between the disciplines, and in doing so to attend to the identities and affective orientations students bring to the classroom. Our argument in this paper points to access points that can shift a student's outgoing relationship with physics and in turn impact his/her participation in physics, as well as impact the ways s/he sees physics as related to domains they care about. We have shown in this case study that a student's starting place is not immutable, a course can engage with a students' starting relationship in order to both impact their participation in physics and to help them see the disciplines as relevant to each other.

## Acknowledgments

We greatly appreciate the feedback on the development of this work from the University of Maryland PERG and BERG. We are particularly grateful for the substantive discussions with and feedback from Benjamin Dreyfus, Benjamin Geller, Julia Gouvea, Ayush Gupta, Eric Kuo, Andrew Elby, and Abigail Daane. This work was supported by NSF-TUES DUE 11-22818 and the HHMI NEXUS grant.